# PGDM-MRSRGAN: Physics-Guided Degradation Model with an SRGAN Framework for Magnetic Resonance Image Super-Resolution: Applications in Low-Field MRI

**Yashwant Kurmi[1,2], Charlotte R. Sappo[1,2,3,4], Sai Abitha Srinivas[5], Selin Kandemir[1,6], Malvika Viswanathan[1,6], Zhongliang Zu[1,2,6]**

[1]Vanderbilt University Institute of Imaging Science, Vanderbilt University Medical Center, Nashville, US. [2]Department of Radiology and Radiological Sciences, Vanderbilt University Medical Center, Nashville, US. [3]Department of Electrical and Computer Engineering, Vanderbilt University, Nashville, US. [4]Department of Chemical and Biomolecular Engineering, Vanderbilt University, Nashville, TN, US. [5] Department of Biomedical Engineering, Case Western Reserve University, Clevland, OH, US, [6]Department of Biomedical Engineering, Vanderbilt University, Nashville, US.

**Word counts: Total 5383; Abstract 220.**

Correspondence to:

Zhongliang Zu, Ph.D.

Vanderbilt University Institute of Imaging Science

1161 21st Ave. S, Medical Center North, AAA-3112

Nashville, TN 37232-2310

Email: zhongliang.zu@vumc.org

Phone: 615-875-9815

Fax: 615-322-0734

Grant Sponsor: R01 EB029443

Running title: **Physics-Guided Degradation Model with an SRGAN Framework for Magnetic Resonance Imaging.**

**ABSTRACT**

**Purpose:** Magnetic Resonance Imaging (MRI) often suffers from low signal-to-noise ratio (SNR) and limited spatial resolution, which compromise clinical precision. This study aims to address these challenges by developing a physics-guided degradation model (PGDM) and a novel deep learning framework, Magnetic Resonance Super-Resolution GAN Imaging (MRSRGAN), for improved super-resolution reconstruction of MRI images.

**Method:** The proposed approach consists of two stages: (1) a physics-guided degradation model that simulates low-field MRI conditions by incorporating artifacts such as blur, B0 inhomogeneity, chemical shift effects, down-sampling, and noise to generate paired low-resolution (LR) and high-resolution (HR) training datasets; and (2) an MRSRGAN-based reconstruction network utilizing a hybrid loss function, including modified spectral angle mapper (MSAM) and sharpened ground truth images based on de-convolution of slice profile-based kernels to restore high-SNR and high-resolution images.

**Results:** Experimental validation demonstrates that the proposed method achieves superior spatial resolution, enhanced SNR, improved peak signal-to-noise ratio (PSNR), higher structural similarity index (SSIM), reduced MSAM error, and better non-reference image quality evaluation (NIQE) metrics. Furthermore, it effectively reduces artifacts and demonstrates robustness on real MRI datasets.

**Conclusion:** The MRSRGAN framework, guided by the physics-informed degradation model, provides a significant improvement in MRI image quality, enhancing spatial resolution and diagnostic accuracy. Its demonstrated robustness and effectiveness on real MRI datasets highlight its potential to resolution enhancement for precise diagnostic imaging.

## 1. INTRODUCTION

Magnetic resonance imaging (MRI) is a non-invasive imaging modality widely used for clinical diagnosis and research due to its ability to provide detailed structural, functional, and molecular insights into soft tissue anatomy without the risks of radiation exposure (1). High-resolution MRI is essential for accurate diagnosis and image processing, providing detailed anatomical insights. Low-field MRI systems (<1 Tesla) face challenges like reduced signal-to-noise ratio (SNR), leading to lower spatial resolution and longer scan times (2-6). These limitations affect diagnostic quality, especially in resource-limited settings or for patients with metallic implants incompatible with high-field MRI systems, underscoring the need for advance innovative solutions (7).

Efforts to mitigate these challenges traditionally rely on hardware upgrades, such as advanced gradient coils or multi-channel receiver systems, which are often cost-prohibitive and impractical in certain settings (6). Alternatively, computational post-processing techniques, including denoising algorithms and interpolation-based super-resolution methods, provide a scalable and cost-effective solution for optimizing image quality. However, these methods often fail to preserve fine structural details, limiting their utility for complex clinical cases requiring detailed anatomical visualization (8, 9).

Deep learning has recently emerged as a powerful tool for improving MRI image quality (10). Generative adversarial networks (GANs), a subset of deep learning models, excel in super-resolution tasks, particularly in medical imaging, by reconstructing high-resolution images from degraded inputs (11). These architectures enhance SNR, spatial resolution, and artifact suppression. However, traditional GAN models often use synthetic training datasets derived from optical degradation models (12) which fail to capture MRI-specific properties (9). These models struggle to accurately simulate MRI-specific artifacts (13), such as isotropic and anisotropic blurring, $B_0$ field inhomogeneity and susceptibility (14), chemical shift effects, and noise, leading to a performance gap when applied to real-world MRI data (15-18).

To address the limitations of low-field MRI systems, we propose a physics-guided degradation model integrated with a deep learning framework, termed Magnetic Resonance Super-Resolution GAN Imaging (MRSRGAN). This method enhances SNR and spatial resolution, building on Srinivas's prior work on electromagnetic interference (EMI) mitigation via the EDITER method (19). The proposed approach incorporates a realistic degradation pipeline to simulate low-field MRI artifacts for training. The reconstruction network, based on a GAN architecture, restores high-SNR and high-resolution images. The degradation model addresses blur, noise, distortions, and chemical shift effects, while the MRSRGAN employs advanced dense blocks and sensitive activation functions, ensuring artifact suppression and structural detail preservation.

Our framework employs novel spectral angle mapper that is a modified version of spectral angle mapper termed MSAM loss using angle image from Fourier-transformed real and imaginary components, mimicking the MRI formation process for optimized loss (20). The sharpened ground truth images based on de-convolution of slice profile-based kernel enhances generalization across diverse MRI modalities and conditions, sharpening high-frequency ground truth images via slice profile de-convolution.

## 2. METHOD

The This study introduces two components for improved MRI reconstruction: 1) a PGDM (Fig. 1A), 2) the MRSRGAN (Fig. 1B, C) as a generator-discriminator.

### 2.1. MRI Physics-Guided Degradation Model

A pipeline is developed to simulate real-world MRI artifacts, incorporating random application of three artifacts from 11 types (Table-I), such as motion artifacts, Gibbs ringing, $B_0$-inhomogeneity, and distortions. These artifacts are sequentially applied to a high-resolution MRI image $I_{HR}$ resulting in a low-resolution image $I_{LR}$ that mimics clinically relevant degradation scenarios. Importantly, only three artifacts are applied per image to maintain tractability of the reconstruction problem. The random selection of artifacts ensures diversity and robustness, enhancing the pipeline's ability to represent realistic MRI conditions. The sequential application of artifacts is mathematically expressed as:

$$I_{LR} = \left[ \left\{ \left\{ A_{i \neq j \neq k} \left( A_{j \neq i \neq k} \left( A_{k \neq i \neq j} (I_{HR}) \right) \right) \right\} * K \right\}_{\downarrow s} \right] + N \qquad (1)$$

$\{where\ \ i = j = k = 1,2, \dots ,11\}$ and $(i, j, k)$ are indices correspond to the selected artifact types, ensuring no repetition of artifacts within each degraded image. After the sequential application of artifacts, a Gaussian blur kernel (K) is applied to further degrade the image. The kernel is spatial domain is defined as $K = \frac{1}{2\pi\sigma^2} e^{-\frac{x^2+y^2}{2\sigma_I^2}}$ where $(\sigma_I)$ is the standard deviation that controls the intensity of the blur. This step mimics resolution loss typically introduced by imaging systems. Following the blurring process, the image is downsampled by a factor (s) and the noise (N) addition can be Rician noise $((N_R))$ and JPEG compression noise.

#### 2.1.1. Artifacts Degradation

2.1.1.1.***Blurring effects*****:** Blurring in MRI is due to several factors.

**Slice thickness effect ($A_1$):** Slice profile thickness-induced blurring in MRI is modeled by averaging consecutive images, accurately representing degradation due to partial volume effects. (21). The average image for each batch is calculated as $I_b = \frac{1}{N_b}\sum_{i=1}^{N_b}(I_i)$. where $N_b$ [3 to 5] is the number of images.

**Limited Frequency Encoding Blur($A_2$):** A 1D Gaussian kernel convolved with the image in the frequency direction $(x)$ as image $I_{slice}(x,y) = I(x,y) * \frac{1}{\sqrt{2\pi\sigma_F^2}} e^{-(\frac{x^2}{2\sigma_F^2})}$. Here, $\sigma_F$ is the standard deviation of point spread along frequency encoding direction, controlling the extent of blur (22, 23).

**Limited Phase Encoding Blur($A_3$):** Limited phase encoding leads to resolution degradation by restricting k-space data along the phase encoding direction (22). For Fourier transform $F(u,v)$ of an image, the degraded image is given by: $I_{phase}(x,y) = F^{-1}(F_{limited}(u,v))$ ; with $F_{limited}(u,v) = \begin{cases} \mathrm{F(u,v)}; & for \quad u \in [\mathrm{u}_{start}: stepSize: u_{end}] \\ 0; & else \end{cases}$. Where $F^{-1}$ represents the inverse Fourier transform, and the restricted k-space data is defined by an arbitrary start frequency $\mathrm{u}_{start}$ end frequency $u_{end}$ and step size within this range.

**Motion Artifacts:** The motion artifacts where subject movement during scanning can result in blurring (**$A_4$**), ghosting (**$A_5$**) or misregistration, adversely affecting image quality (22). Motion artifacts are simulated in the frequency domain (k-space) by applying motion-induced phase shifts: $k_{motion-blur}(u,v) = k(u,v).e^{j2\pi f_{motion}\,y}$; where $k(u,v)$ represents the original k-space data, $f_{motion}$ is the normalized motion frequency, and $e^{j2\pi f_{motion}\,y}$ introduces periodic phase shifts along the phase-encoding direction (24).

2.1.1.2.**Ghosting effect** is added by modulating the contributions of specific phase-encoding lines (22):

$$k_{mg}(:,v) = \begin{cases} k_m(:,v).g_F; & if\ v \in \{0,2,4,\dots\} \\ k_m(:,v) & if\ v \in \{1,3,5,\dots\} \end{cases} \tag{2}$$

Where $k_{mg}$ is motion-ghost effect or degraded k-space data for $k_m$ motion factor and it is transformed back to the spatial domain using the inverse Fourier transform: $I_{mg}(x,y) = \left|F^{-1}\left(k_{mg}(u,v)\right)\right|$ and $g_F: ghost - Factor$ depends on periodic motion (breathing, cardiac) frequency.

**2.1.1.3. Chemical shift artifacts ($A_6$):** Chemical shift artifacts in MRI arise from differences in resonance frequencies between fat and water molecules, leading to spatial misalignment, edge shifts, or blurring (25). Artifacts are simulated by applying frequency-dependent distortions in k-space, introducing phase shifts or pixel displacements along the frequency encoding direction to the fat component. The frequency shift is computed by $f_{\text{Chemical}_{\text{Shift}}}(Hz) = f_{\text{fat}_{\text{freq}_{\text{shift}}}} \times \gamma \times B_0$, with chemical shift $f_{\text{fat_freq_shift}}$ is in ppm, $\gamma$ is the gyromagnetic ratio (42.58 (MHz/T)), and $B_0$ is the magnetic field strength in Tesla. The frequency shift in pixels is derived as: $f_{\text{fat_freq_shift}} = f_{\text{Chemical}_{\text{Shift}}}(Hz)/\text{bandwidth}$ (Table-I). This shift is applied to the fat k-space data, causing a spatial misregistration along the frequency encoding direction and added to the water k-space data. The combined water and shifted fat signals are reconstructed using inverse Fourier transform.

2.1.1.4. **Ringing Artifacts ($A_7$):** Gibbs ringing artifacts arise from k-space truncation or undersampling, introducing high-frequency oscillations near sharp edges in the spatial domain (26, 27). The reconstructed spatial signal due to truncation is expressed as:

$$f_{trunc}(x) = \int_{-k_{max}}^{k_{max}} \mathrm{F(k)}\ e^{j2\pi kx}\, dk \quad (3)$$

This restriction in k-space leads to oscillatory patterns, commonly known as Gibbs ringing artifacts. Additionally, sinc interpolation, defined as $sinc(x) = sin(\pi x)/\pi x$ typically used for resampling MRI data, can represent these artifacts due to the oscillatory nature of the sinc kernel.

2.1.1.5. **Field Inhomogeneity ($A_8$):** Field inhomogeneity, which causes spatial and intensity distortions in MRI, is modeled to evaluate its effect on image reconstruction (25, 28). The ideal k-space data, $\boldsymbol{k}_{ideal}$, is obtained from the Fourier transform of image $\boldsymbol{x}$ as: $\boldsymbol{k}_{ideal} = FFT2(\boldsymbol{x})$. A Gaussian inhomogeneity map, $\boldsymbol{H}(x,y) = e^{-\left(\frac{x^2+y^2}{2\sigma_{\Delta B_0}^2}\right)}$ scaled by $\alpha$ ∈[0.05-0.3] as $\boldsymbol{H}_{scaled} = \alpha.\ \boldsymbol{H}(x,y)$, introduces distortions. The distorted k-space is: $\boldsymbol{k}_{distorted} = \boldsymbol{k}_{ideal}.\, e^{(j2\pi \boldsymbol{H}_{scaled})}$. Reconstructed image was obtained via the inverse Fourier transform: $\boldsymbol{x}_{distorted} = \left(IFFT2(\boldsymbol{k}_{distorted})\right)$. Using combination of multiple Gaussian models with varying radius helps create greater variations, enabling the generation of more realistic inhomogeneity artifact in image.

2.1.1.6. **Signal Void and Shading Artifact($A_9$):** The shading artifacts in MRI images were simulated (29) using a Rician kernel derived from the Rician PDF:

$$f(r|\vartheta, \sigma_R) = \frac{r}{\sigma_R^2} \exp\left(\frac{-(r^2+\vartheta^2)}{2\sigma_R^2}\right) I_0 \left(\frac{r\vartheta}{\sigma_R^2}\right) \quad (4)$$

Where $r \in \left[\frac{imageSize}{2} \geq 0\right], \sigma_R = 1$, and $\vartheta = 0.53$. The kernel was normalized and applied in both frequency (via convolution) and spatial domains (via filtering) to mimic the shading effect. First, it applied in K-space, and the degraded K-space data was transformed back to the spatial domain using the inverse Fourier Transform, followed by direct spatial filtering.

2.1.1.7. **Magnetic Susceptibility (A10):** Magnetic susceptibility artifacts in MRI images were simulated (14, 30, 31) based on the spin density ($\rho$) and voxel volume ($\Delta R$) at a spatial position ($r$): $I(r) = \rho(r)\Delta R$. Off-resonance frequencies ($\Delta$) induce distortions, causing bulk shifts along the frequency-encoding direction, expressed as changes in spatial encoding. Additionally, local gradients of off-resonance frequencies $\left(\frac{\partial \delta v(r)}{\partial r}\right)$ impact voxel volumes $\delta v(r)$, causing localized hyper-intensities or hypo-intensities. The artifacts were simulated (30, 31) by detecting edges in the image using the Sobel operator. A random edge location served as the center for a spherical high/low-intensity kernel ($I_{spherical_kernel}$) constructed using a Gaussian function. The kernel was applied to the image ($I_{Original}$), introducing localized distortions. The artifact variations were generated using logarithmic transformations: $I_{\text{Susceptibility}} = log(1 + I_{Original} + I_{spherical_kernel})$ (32, 33) along with additional models and parameters from (29).

2.1.1.8. **Zebra Artifacts (A11):** A sinusoidal $\Delta B_0$ field distortion is introduced, defined as $\Delta B_0(x, y) = A \,.\, sin\left(2\pi f . \frac{x}{N}\right)$. where $A$ is the distortion amplitude, $f$ is the sinusoidal frequency, and $N$ is the image size. This distortion creates a periodic variation across the spatial domain of the MRI image (13). The Fourier transform, $K_{original}(u, v) = F\big(I_{normalized}(x, y)\big)$ is applied to the normalized MRI image to obtain k-space data. Using the $\Delta B_0$ distortion map, a phase modulation map is generated $K_{phase}(u, v) = \mathrm{e}^{(2\pi i . \Delta B_0)}$ and applied to the k-space data to introduce the artifact. The distorted k-space data is then transformed back to the spatial domain using an inverse Fourier transform, $I_{\text{artifact}}(x, y) = F^{-1}\big(K_{original}(u, v). \mathrm{e}^{(2\pi i . \Delta B_0)}\big)$ to reconstruct the artifact-affected image.

### 2.1.2. Down Sampling (D1)

Downsampling is essential for super-resolution MRI preprocessing, using methods like row-only, column-only, or both dimensions (34). Bicubic interpolation ensures spatial consistency, while normalization maintains intensity scaling across different degradation scales (35).

### 2.1.3. Noise Degradation ($N_1$)

Noise degradation is simulated by adding Rician and Gaussian noise to high-resolution images after incorporating typical MRI artifacts. Rician noise is generated by scaling the signal with the Rician (K)-factor and adding random noise, expressed as:

$$R = \sqrt[2]{(s + n_{real})^2 + (n_{imag})^2} \quad (5)$$

Where $s$ is the scaled signal, $n_{real}$ and $n_{imag}$ are the Gaussian noise with standard deviation ($\sigma = \sqrt[2]{\frac{1}{2(K+1)}}$. Gaussian noise is modeled as: $G = \sigma_G . N(0,1)$. Where $\sigma_G$ is the standard deviation and $N(0,1)$ is a zero-mean unit variance Gaussian distribution. The final noisy image is obtained by combining Rician and Gaussian noise: $[I_{noisy} = I_{Original} + R + G]$ (36).

### 2.1.4. JPEG compression noise ($N_2$)

JPEG compression uses the discrete cosine transform (DCT) to divide an image into (8×8) blocks and quantize using a quantization matrix. Lossy compression introduces artifacts like blocking and detail loss. The de-compressed image is reconstructed via inverse quantization and inverse-DCT. It simulates real-world compression-induced degradation (37).

## 2.2.Network Architectures and Training

We adopt the base ESRGAN (38) model with modifications, where each RRDB contains three residual blocks. To enhance model sensitivity, the LeakyReLU activation function with a negative slope of 0.25 is used in the generator. The MRSRGAN model includes 19 RRDBs, optimized for advanced MRI reconstruction using a hybrid loss function that incorporates MSAM loss (39) and high frequency sharpened ground truth images based on de-convolution of slice profile. The discriminator employs spectral normalization and LeakyReLU (negative slope 0.25), termed the extreme sensitive discriminator (ESDiscriminator) for enhanced sensitivity.

## 2.3.Loss Functions

The MRSRGAN architecture leverages a hybrid loss function ($l^{SR}$) given by Eq. (8) to optimize the high-resolution image reconstruction includes the following losses:

**2.3.1. Content Loss** ($l_{MSE}^{SR}$): Measures pixel-wise differences between reconstructed and reference images using Mean Squared Error (MSE) as $l_{MSE}^{SR} = \frac{1}{rW \times rH} \sum_{x=1}^{rW} \sum_{y=1}^{rH} \left(I_{x,y}^{HR} - G_{\theta_G}(I^{LR})_{x,y}\right)^2$. Where $I^{HR}$ and $I^{LR}$ represent the high-resolution ground truth and LR input images, respectively, and $G_{\theta_G}(I^{LR})$ is the generator network output.

**2.3.2. Perceptual Loss ($l_{VGG}^{SR}$)**: Captures high-level structural features by comparing feature maps from a pre-trained VGG19 network (40). It emphasizes feature-level similarity between $\phi\left(G_{\theta_G}(I^{LR})\right)$ and $\phi(I^{HR})$:

$$[l_{VGG/j}^{SR} = \frac{1}{W_j H_j} \sum_{x=1}^{W_j} \sum_{y=1}^{H_j} \left(\phi_j(I^{HR})_{x,y} - \phi_j\left(G_{\theta_G}(I^{LR})\right)_{x,y}\right)^2 \quad (6)$$

Here $W_j$ and $H_j$ denote the width and height of feature maps at the $jth$ layer of VGG19, respectively, while $\phi_j$ represents the feature extraction operations.

**2.3.3. Adversarial Loss ($l_{Adv}^{SR}$):** Encourages the generator to produce realistic images that can fool the discriminator. It is computed as: $l_{Adv}^{SR} = \sum_{n=1}^{N} -log D_{\theta_D}\left(G_{\theta_G}(I^{LR})\right)$ with discriminator $D_{\theta_D}$ and generated image $G_{\theta_G}(I^{LR})$.

**2.3.4. Modified Spectral Angle Mapper (MSAM) Loss ($l_{MSAM}^{SR}$)**: Evaluates spectral consistency between the high-resolution ground truth and the reconstructed image. It uses the 2D Fourier transform, frequency shifting, and normalization to calculate the mean squared error of the absolute phase angle images $I_A = arctan(Img(I)/Re(I))$ preserving k-space features (39).

$$l_{MSAM}^{SR} = \frac{1}{rW \times rH} \sum_{x=1}^{rW} \sum_{y=1}^{rH} \left({I_{x,y}^{HR}}_A - G_{\theta_G}(I^{LR})_{{x,y}_A}\right)^2 \quad (7)$$

The hybrid loss ensures structural accuracy, visual realism, and phase angle image by combining multiple loss components:

$$l^{SR} = l_{MSAM}^{SR} + l_{MSE}^{SR} + l_{VGG}^{SR} + 10^{-2}\, l_{Adv}^{SR} \quad (8)$$

### 2.4.Training strategy

The training process is divided into three stages. First, the generator model is trained using a dataset of MRI images with artifacts and downgraded quality aiming to produce stable and high-quality images. PSNR is improved using content-loss during this step. Next, the generator is pretrained within the MRSRGAN framework, employing a hybrid loss function that combines content-loss, perceptual loss (41), GAN loss, and the proposed MSAM loss. High-frequency image components are enhanced through deconvolution of slice-profiles, commonly found in MRI data scanning (42). Finally, the pretrained model undergoes fine-tuning on a dataset degraded by specific artifact types. Finally, the pretrained model undergoes fine-tuning on a dataset degraded by specific artifact types. This fine-tuning ensures the model is optimized for accurate prediction.

To simulate realistic low-resolution MRI images, a random shuffle strategy reorders degradation processes, including artifact simulation, downsampling, noise addition, and JPEG compression. Downsampling employs bicubic methods with randomly selected scale factors, ensuring adaptability and enhancing training robustness.

### 2.5. Performance parameters

The performance parameters include SNR, PSNR, Structural similarity index (SSIM), along with the multi-scale structural similarity index (MSSSIM) (43), noise quality measure (NQM) (44). The NIQE (45, 46) evaluate reconstructed image quality by measuring naturalness and minimizing artifacts. Higher SNR, PSNR, SSIM, NQM, and MSSSIM values indicate better performance, while lower MSAM and NIQE values signify improved reconstruction.

## 3. Results

### 3.1. Datasets and Implementation

**3.1.1. Synthetic Data:** The M4Raw dataset (47) used for the MRSRGAN framework comprises $T_2$-weighted $T_1$-weighted and FLAIR MRI images at 0.3 Tesla. It contains high-resolution (HR) images of size 256, with HR patches also set to 256. training was conducted on an NVIDIA RTX A4000 GPU with a batch size of 12, using the Adam optimizer (48). The generator was pre-trained for 500K iterations with a learning rate of $2\times10^{-4}$ to produce meaningful outputs for discriminator training. Subsequently, MRSRGAN was trained for 500K iterations with learning rate $1\times10^{-4}$ employing bias-corrected exponential moving average (49). The model optimizes a combination of content-loss, perceptual loss, GAN loss, and MSAM loss, with respective weights {1, 0.1, 0.1, 0.1}. For the perceptual loss, we utilize feature maps from the first five convolutional layers of the pretrained VGG19 network(40), with corresponding weights of {0.1, 0.2, 0.4, 0.8, 1}, applied before the activation function. Our implementation is built upon the BasicSR framework (50). Parameters of MRSRGAN model are 18M with 19 RRDB blocks.

**3.1.2. In Vivo Data:** All in-vivo data were acquired on a 47.5 mT biplanar permanent magnet MRI system (Sigwa MRI, Boston, MA, USA) (19). The open biplanar magnet has a main magnetic field ($B_0$) that provides good homogeneity over the imaging volume to enable hard pulse RF excitation, as described in (19). The scanner is equipped with three-axis planar gradient coils and is controlled using a Tecmag Redstone console (Tecmag, Houston, TX, USA) with two transmit and three receive channels. Gradient waveforms are each driven by AE Techron 2120 amplifiers (Elkhart, IN, USA), and RF transmission is performed using a 2-kW peak-power amplifier (Tomco Technologies model BT02000-AlphaS, Stepney, Australia). Signal acquisition was performed using a single channel transmit/receive RF coil (6, 19, 51, 52), with a secondary (non-imaging) coil and electrode for EMI sensing and removal as described in (19).

In vivo imaging experiments were conducted in ten healthy volunteers following written informed consent, in accordance with the Vanderbilt University Medical Center Institutional Review Board (IRB) guidelines. All imaging data were acquired using a three-dimensional (3D) multi-echo Rapid Acquisition with Relaxation Enhancement (RARE) spin-echo sequence (51-53) as further described in (19). The sequence parameters were: spatial resolution = $1 \times 2 \times 9$ mm$^3$, acquisition matrix = 128 (readout) × 97 (in-plane phase encode) × 23 (partition encode). To evaluate the robustness of the proposed method to external noise, data were acquired both with and without EMI mitigation using the EDITER method (19).

**3.1.3. Degradation details:** The first order degradation is based on the MRI artifacts in Table-I, while the second-order degradation model enhances effectiveness by simulating typical degradation processes. Gaussian, generalized Gaussian, and plateau-shaped kernels are used with probabilities {0.6, 0.2, 0.2}, respectively. The shape parameter ($\beta$) is sampled from [0.5, 2] for generalized Gaussian kernels and [1, 2] for plateau-shaped kernels. A sinc kernel is incorporated with a probability of 0.1. The second blur degradation step is skipped with a probability of 0.5.

We employ Gaussian noise and Rician noise, each with a probability of 0.5. The noise sigma range for Gaussian and Rician noise is set to [1, 25] and [0, 10], respectively.

### 3.2. Training Data pairs

To optimize training efficiency, all degradation processes are implemented in PyTorch with CUDA acceleration. In each iteration, training samples are randomly drawn from the set of degraded pairs to form a training batch.

**3.2.1. Sharpen ground-truth images during training:** We also introduce a fine-tuning technique with enhanced visual sharpness reference images without causing noticeable artifacts. An approach to sharpening images involves slice profile deconvolution, which accounts for the slice profile thickness used in MR image scanning. The deconvolution of slice profile of range [2-5] was used in image sharpening (42).

**3.2.2. Comparisons with prior works:** We compare the proposed MRSRGAN method with SRGAN(12), ESRGAN(54), BSRGAN(55), and Real-ESRGAN(38). Notably, since our contribution lies in the degradation model, SRGAN, ESRGAN, BSRGAN, and Real-ESRGAN share the same network architecture as ours, we did not re-train these models for comparison.

**3.2.3. Testing Datasets**: The MR-ART dataset, comprising T1-weighted 3D MRI images under varying motion conditions, was used to assess motion-related artifact correction methods (56). Synthesized degradation processes simulate low-field MRI artifacts like chemical shift artifacts, $B_0$-inhomogeneity, and noise.

**3.2.4. Testing Datasets**: To facilitate the evaluation of motion-related artefact correction methods, we utilized the MR-ART dataset, which consists of structural T1-weighted 3D MRI images collected under varying motion conditions. The dataset provides matched motion-free and motion-corrupted images, enabling direct analysis of motion artefacts and their impact. Additionally, the dataset includes clinical artefact scores from neuroradiologists and standardized image quality metrics, offering a comprehensive resource for testing and improving motion correction approaches (56). Other degradation processes have been synthesized to account for common artifacts observed in low-field MRI, including $B_0$ inhomogeneity artifacts, chemical shift artifacts, and noise-related degradation. These artifact reduction methods are demonstrated on MRI images, alongside resolution enhancement, to showcase improvements in image quality and address the challenges posed by these artifacts.

### 3.3. *Intermediate Results on the Proposed MSAM Angle*

Fig. 2 shows the In-phase and Q-phase images comparison of different super resolution methods in MRI image enhancement for SRGAN, BSRGAN, ESRGAN, Real-ESRGAN, MRSRGAN and reference for comparison. It can be seen the less deviation in the outcome of MRSRGAN matches the outcomes of In-phase and Q-phase images from the reference.

### 3.4. *Comparison of Qualitative Results*

**3.4.1. Result comparison using simulated dataset:** Standard Shepp–Logan phantom (SLP) was simulated for general validation of imaging processes (57). A second custom digital phantom (CDP) dataset was created, as shown in Fig. 3, consisting of two distinct frequency components representing the image intensities of water and fat signals separately. Chemical shift effect occurs due to the slight frequency difference between the resonant frequencies of hydrogen nuclei in water and fat molecules.

**3.4.1.1. Case 1:** Fig. 3 demonstrates the effectiveness of the proposed model in removing chemical shift artifacts. In CDP images, medium intensity represents water and high intensity represents fat, while chemical shift images show fat displacement toward water (became brighter than original) can also be verified with residuals.

Compared to other methods, the proposed approach yields the lowest residual artifacts, indicating improved reconstruction, whereas SRGAN suffers from mode collapse and is excluded from fair comparison. Quantitative results in Supporting Information Table S2 further confirm superiority, with MRSRGAN achieving the best performance (PSNR: 30.54 dB, SSIM: 0.7332, NQM: 26.12, MSSSIM: 0.9476, MSAM: 1.5235).

**3.4.2. Result comparison using MR-ART Dataset:** In this section, we present the results of comparing our approach against other methods using the online available real-world MR-ART dataset. The results of this experiment are reported in lower part of Supporting Information Table S3, and they show similar trends of improvements obtained from the SLP Simulated dataset.

**3.4.2.1. Case 2:** Fig. 4 presents the performance of the proposed method under varying motion conditions and imaging views. In the SLP simulated data (upper five rows), motion-induced distortions are effectively reduced, as evidenced by the residual maps (rows 3 and 5), where the proposed method yields the lowest residuals. Similar trends are observed in the MR-ART dataset (lower five rows), further validating the robustness of artifact detection and removal. With increasing motion severity (Motion-1,-2), the proposed method consistently achieves improved reconstruction in both simulated and in vivo data, outperforming all comparative methods across all views.

Supporting Information Table S3 provides a quantitative comparison of methods for motion artifact enhancement using a SLP simulated and MR-ART datasets. By utilizing the MR-ART dataset, the proposed MRSRGAN surpasses existing solutions like SRGAN, ESRGAN, BSRGAN, and Real-ESRGAN (38) in key metrics such as PSNR (28.08 dB), SSIM (0.7459), NQM (25.79), MSSSIM (0.9391), and MSAM (1.5211).

**3.4.2.2. Case 3:** Fig. 5 illustrates the performance of the proposed method under varying $B_0$-inhomogeneity conditions and views. In both SLP simulated and MR-ART synthetic (artifacts are simulated in the GT images of MR-ART dataset) datasets, it effectively reduces intensity distortions, as reflected by minimal residuals. With increasing inhomogeneity severity ($B_0$-inhomogeneity-1 to 2), MRSRGAN improves reconstruction quality and artifact correction, consistently outperforming other methods and demonstrating robust, efficient handling of $B_0$-related artifacts across all images.

Supporting Information Table S4 supports these findings by presenting a performance comparison of various methods for $B_0$ inhomogeneity artifact-related image enhancement, validated using a simulated dataset.

**3.4.2.3. Case 4:** Fig. 6 highlights the proposed MRSRGAN's effectiveness in identifying artifacts under varying noise levels. Increasing noise (from level 1 to 2) in SLP simulated and MR-ART synthetic data, the MRSRGAN outperformed on alterations with least residuals. Supported by Supporting Information Table S5's the MRSRGAN consistently delivered top performance across all images, in comparison to all state-of-the-art methods.

The MRSRGAN framework was validated using the MR-ART dataset by comparing reconstructed images against high-resolution reference images. Figs 4-6 highlight how specific features aid in identifying various artifacts, facilitating targeted improvements in artifact detection systems.

Performance metrics such as PSNR, SSIM, and MSAM were evaluated, with Motion-1 representing low-level motion and Motion-2 depicting high-level motion as shown in Fig. 7 for 40 samples. The results demonstrate that the proposed MRSRGAN method outperforms all state-of-the-art methods.

### 3.5. Experiments on our LF MRI Dataset

The in-vivo dataset at 47.5 mT low-field MRI was used to assess the performance of the proposed MRSRGAN method, as shown in Figs. 8 and 9. Both figures include zoomed-in versions of all four images, highlighted with red rectangular boxes in the corresponding rows beneath the full image samples. The acquired low-resolution input images were first processed using the EDITER method for EMI removal, as illustrated in Figs. 8(B) and 9(B).

These EDITER-processed images were then used to enhance resolution and overall image quality through various super-resolution methods, as shown in Figs. 8 and 9: (C) SRGAN, (D) ESRGAN, (E) BSRGAN, (F) Real-ESRGAN, and (G) the proposed MRSRGAN method. The visual analysis reveals suboptimal performance from SRGAN and ESRGAN models. SRGAN amplifies artifacts in reconstructed images, while BSRGAN overly smooths textures, eliminating critical structural details despite its focus on noise removal. Real-ESRGAN introduces excessive graininess and hallucinated textures, as seen in Fig. 9(F), compromising the diagnostic utility of the images. These shortcomings highlight limitations in preserving anatomical accuracy and texture in image reconstruction.

Conversely, the proposed MRSRGAN method excels in reducing noise while preserving structural details, offering significantly enhanced resolution in reconstructed images. It achieves better texture preservation and minimizes noise artifacts. Both qualitative and quantitative validations, through visual and tabular analysis, demonstrate the superiority of MRSRGAN in delivering anatomically accurate and diagnostically useful reconstructed images.

Figs. 10(A, B) present SNR and NIQE values for various methods, highlighting the superior performance of the proposed MRSRGAN, which achieved an SNR of 120 dB for normalized images. Despite SRGAN and ESRGAN having relatively low NIQE values (Fig. 10B), they failed to produce satisfactory visual results (Figs. 8C, 9C). For fair comparison, NIQE were evaluated across top performing methods: BSRGAN (1.3), Real-ESRGAN (1.2), and MRSRGAN, which achieved the best NIQE value of 1.1.

## 4. Discussion

The MRSRGAN framework represents a transformative advancement in deep learning-based MRI enhancement, specifically targeting the limitations of low-field systems. Unlike conventional approaches, MRSRGAN optimizes the critical trade-offs between resolution, signal-to-noise ratio (SNR), and acquisition time, a longstanding challenge in the field. While earlier frameworks like SRGAN laid the groundwork for super-resolution techniques, subsequent developments, such as ESRGAN, BSRGAN, and Real-ESRGAN, introduced improvements in perceptual loss and robustness. However, these methods lack domain-specific optimizations, particularly for spectral fidelity in MRI. MRSRGAN addresses this gap through MSAM loss, ensuring more precise spectral reconstruction and realistic super-resolution tailored to MRI applications. MRSRGAN's success in low-field MRI stems from its robust framework, which integrates MSAM loss to preserve spectral relationships and reconstruct texture and intensity variations with precision. Furthermore, its slice-profile deconvolution enhances high-frequency components, improving training quality and image fidelity. The ability to recover absolute and complex image values positions MRSRGAN as a practical tool for diagnostic imaging in resource-constrained environments (59).

The persistent challenge posed by Plenge (60) in 2012, regarding the balance between resolution, SNR, and acquisition time remains a focal point for MRI enhancement research. Despite advancements, performance analysis shows this challenge remains partially unresolved. Despite advancements, low-field imaging systems still grapple with constraints like limited training datasets and hardware inefficiencies. A key innovation lies in MRSRGAN's ability to leverage k-space data, capturing raw frequency and spatial information, which significantly improves reconstruction accuracy over image-space methods. This approach proves particularly effective for low-field imaging scenarios (e.g., 47.5mT), demonstrating that models trained on higher-field datasets, such as 0.3T, can generate high-quality images under resource-limited conditions. By overcoming hardware limitations, MRSRGAN enhances SNR, resolution, and overall imaging quality, making it a viable solution for real-world applications in low-field MRI.

However, the model's broader applicability relies on addressing challenges such as dataset diversity and generalizability. Future research should focus on incorporating domain adaptation techniques to strengthen robustness across varied populations and imaging scenarios. Additionally, integrating deep learning-based EMI mitigation could further refine image quality and reliability, ensuring MRSRGAN's continued advancement in clinical and research applications.

## 5. Conclusion

The physics-guided degradation model-based MRSRGAN framework presents a robust solution for enhancing SNR and spatial resolution in low-field MRI images. By incorporating MRI-specific artifacts, multi-scale down-sampling, and noise into the degradation model, along with advanced generator-discriminator architectures, optimized training dynamics, modified SAM loss, extra-sensitive activation functions, and slice profile deconvolution-based image sharpening, the framework achieves significant improvements across key metrics, including SNR, PSNR, SSIM,

NQM, MSSSIM, MSAM, and NIQE. Validation on simulated and in vivo datasets confirms its reliability, advancing low-field MRI for faster, accurate, and clinically impactful image resolution.

**DATA AVAILABILITY STATEMENT**

All source code for processing and analysis, along with the associated data, will be available as per the appropriate request.

**List of Figures Captions:**

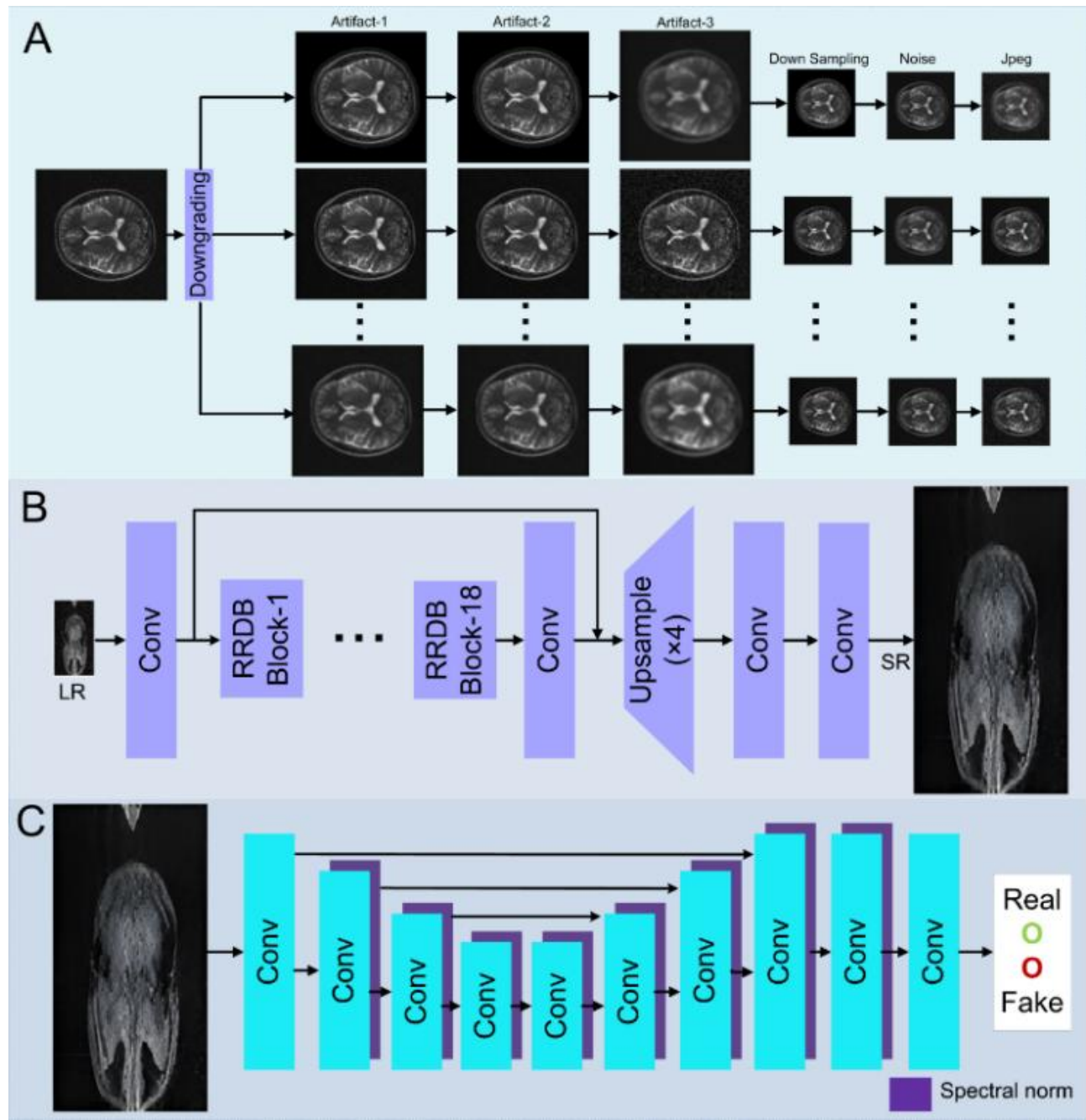

Fig. 1: (A) Illustration of the degradation model incorporating random three MRI artifacts Artifact-1, Artifact-2, and Artifact-3 (out of 11 MRI artifacts), followed by downsampling, noise addition, and JPEG compression, (B) Block diagram of the MRSRGAN, (C) the diagram of extra-sensitive discriminator.

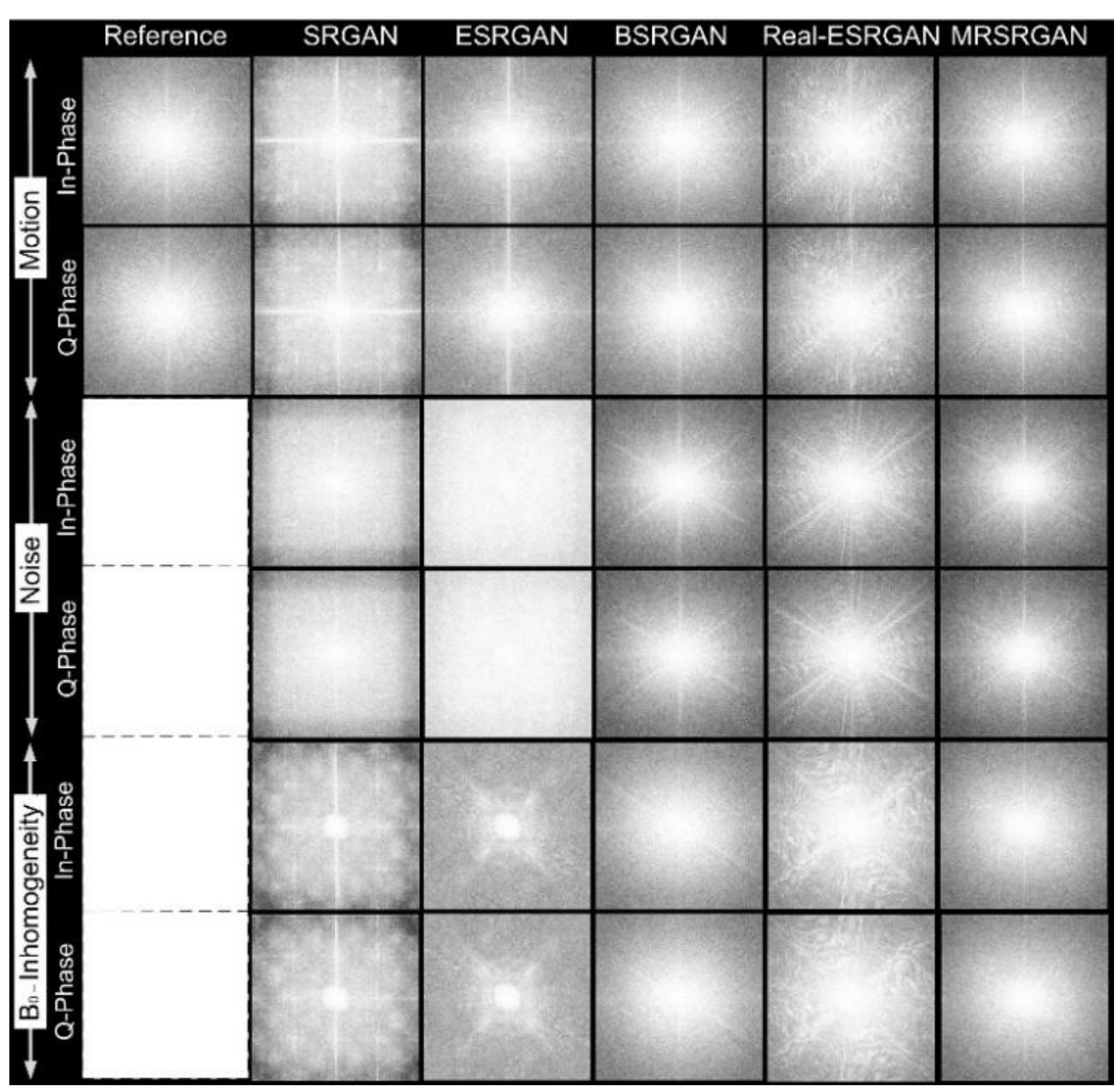

Fig. 2: In-phase and Q-phase images with SRGAN, BSRGAN, ESRGAN, Real-ESRGAN, and MRSRGAN results, showcasing spectral fidelity. The proposed MRSRGAN method outperforms state-of-the-art on the MR-ART dataset.

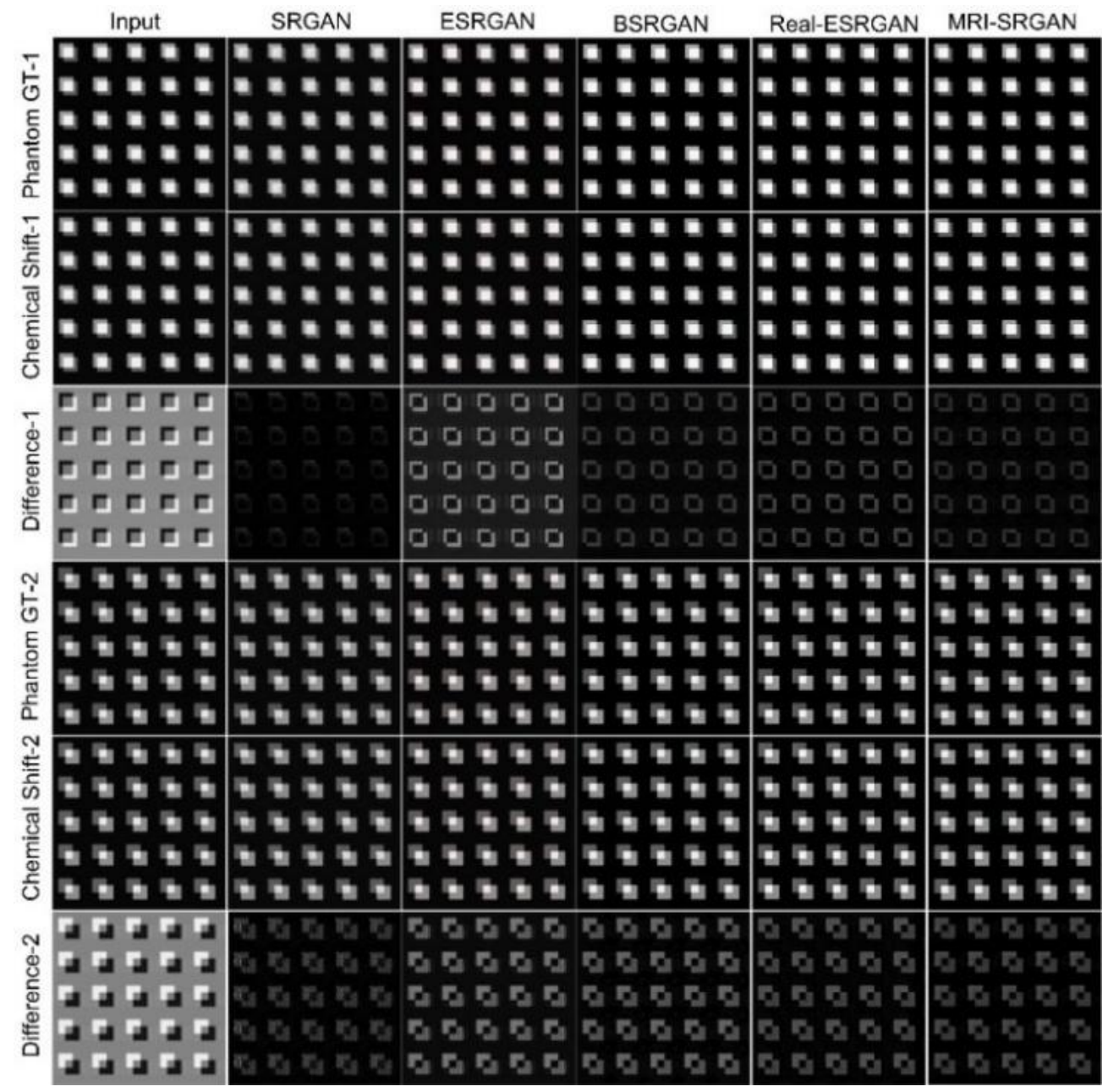

Fig. 3: Illustration of chemical shift artifact (CS) removal using SRGAN, BSRGAN, ESRGAN, Real-ESRGAN, and the proposed MRSRGAN method on two CDP dataset images and outperforms the state-of-the-art approaches.

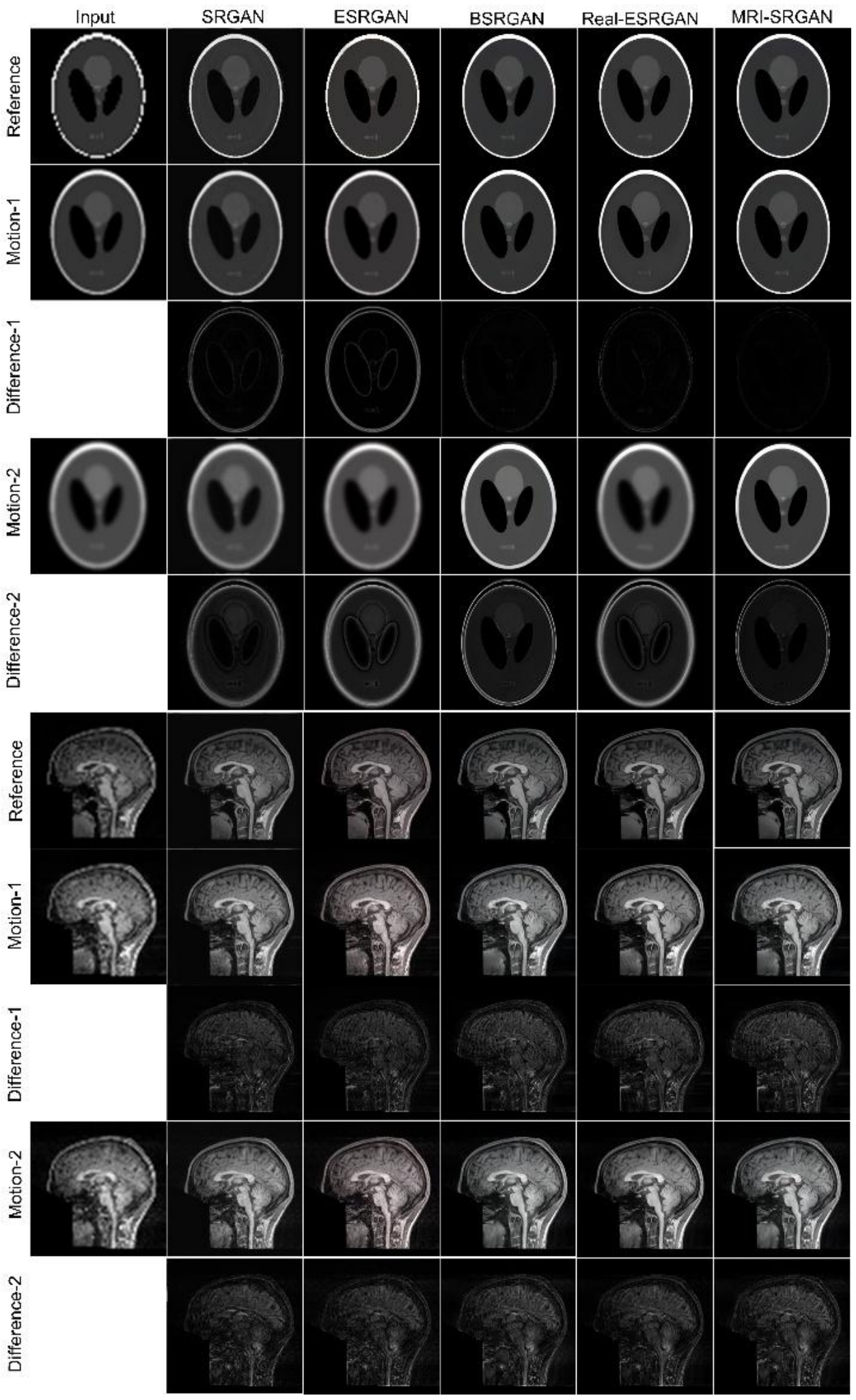

Fig. 4: Comparison of super-resolution methods (top) on simulated SLP data (top 5 rows) and brain images of MR-ART (original) dataset (bottom 5 rows) with motion artifacts, including references and residuals [56, 58]. The proposed MRSRGAN outperforms state of the art methods.

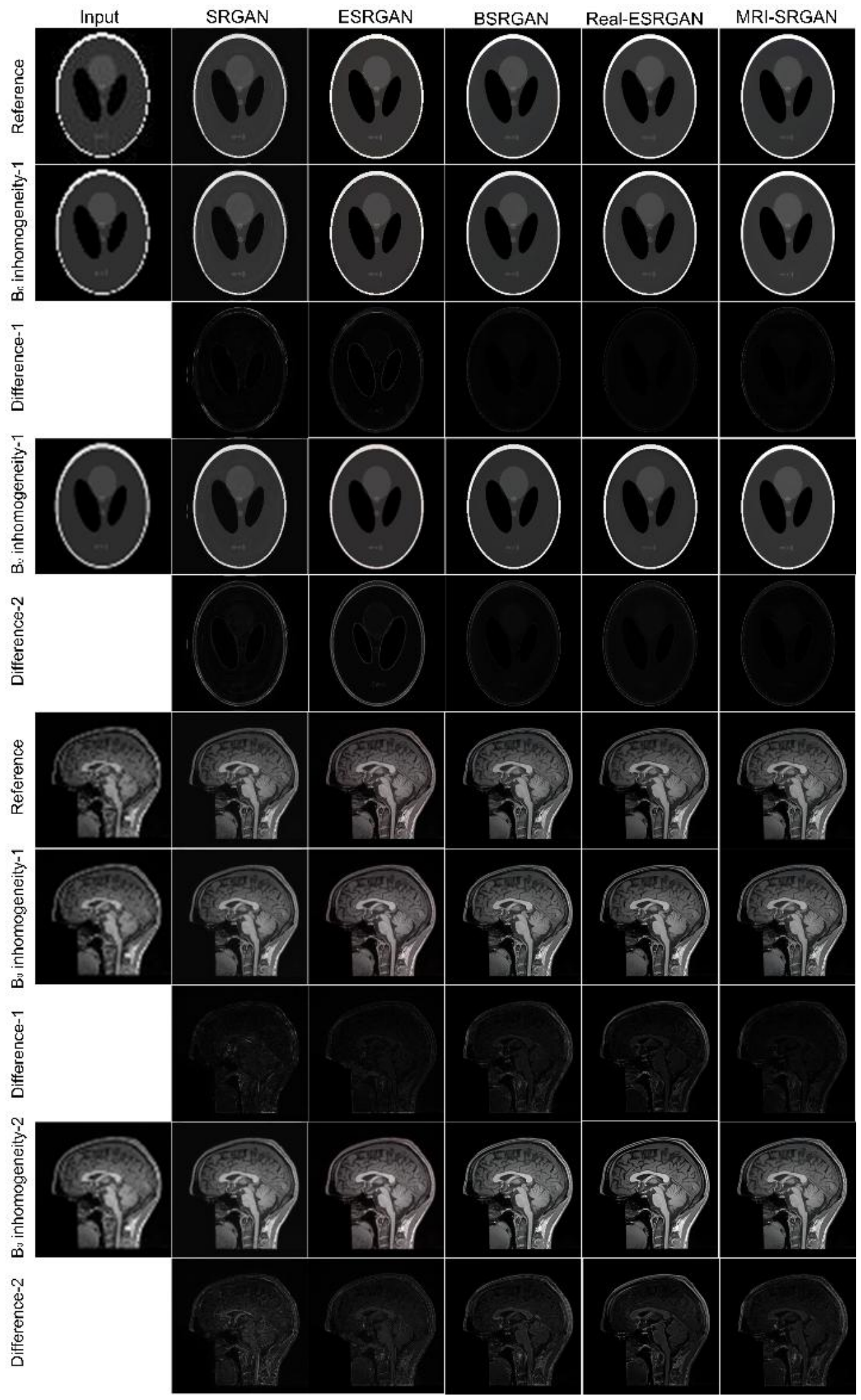

Fig. 5: Comparison of super-resolution methods (top) on simulated SLP data (top 5 rows) and brain images of MR-ART (synthetic) dataset (bottom 5 rows) with B0-inhomogeneity artifact including references and residuals. Overall, the proposed MRSRGAN outperforms state-of-the-art methods.

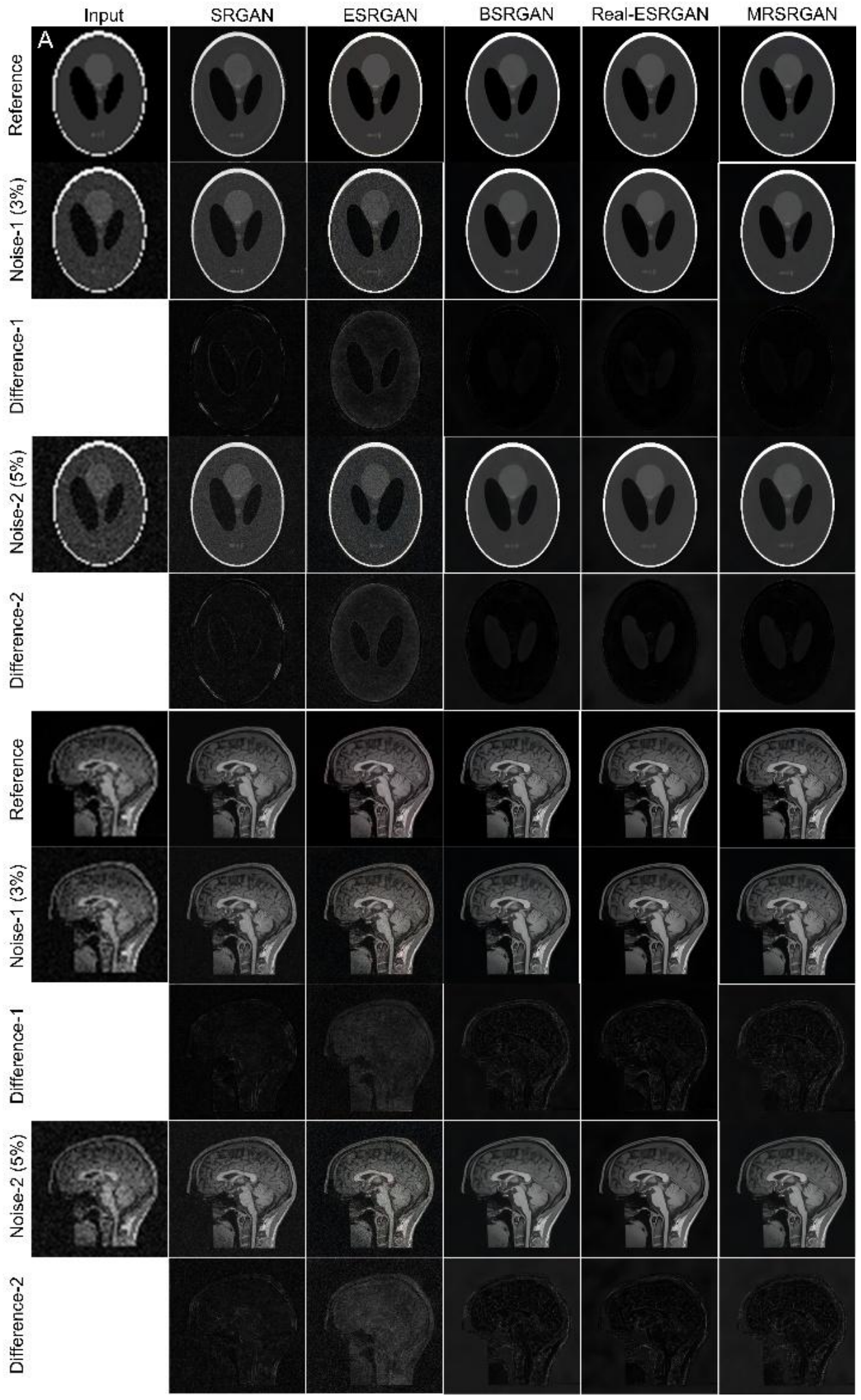

Fig. 6: Comparison of super-resolution methods (top) on simulated SLP phantom data (top 5 rows) and brain images of MR-ART (synthetic) dataset (bottom 5 rows) with Noise (3% &5%) including references and residuals. Overall, the proposed MRSRGAN outperforms state-of-the-art methods.

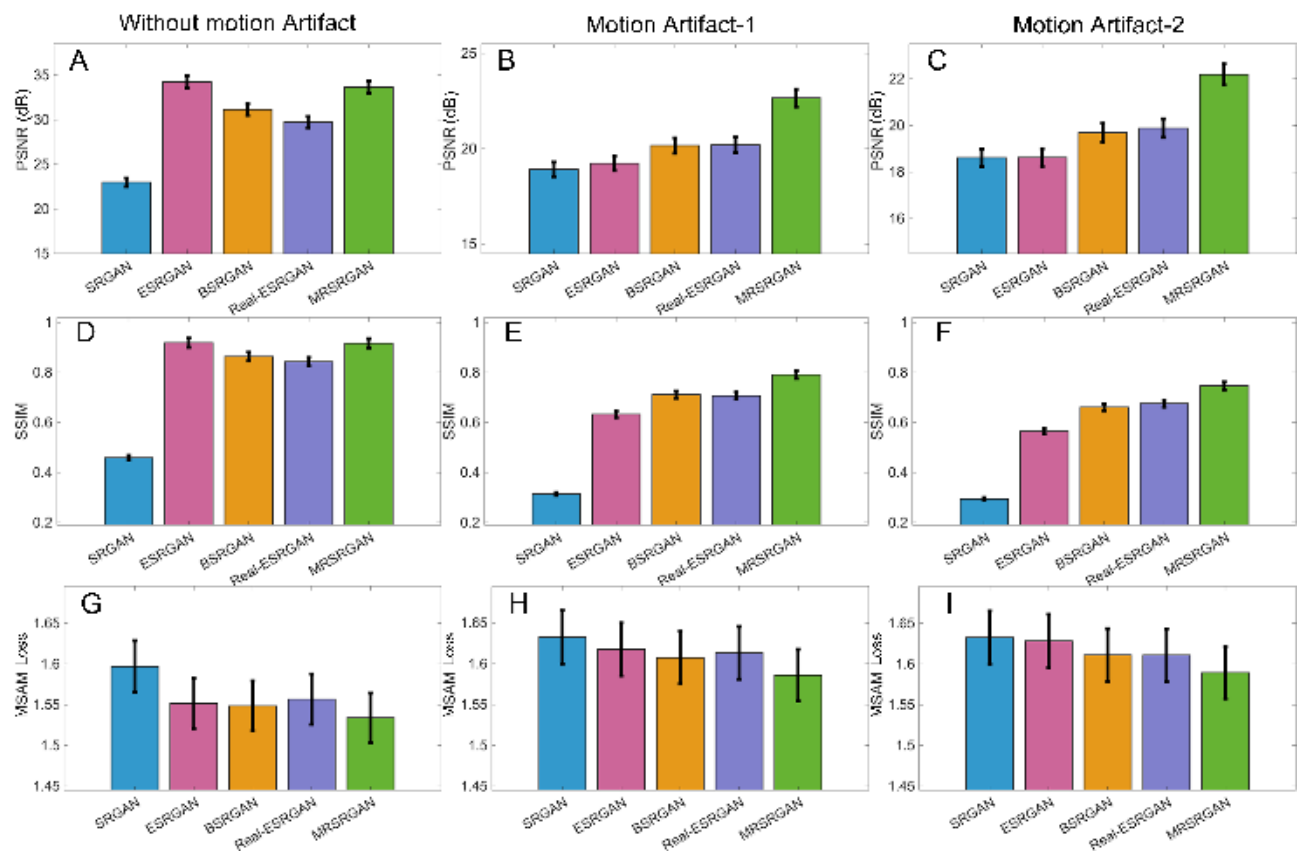

Fig. 7: Performance comparison of various methods on 40 samples from the MR-ART dataset [56, 58]. The performance of the proposed method is superior compared to state-of-the-art methods.

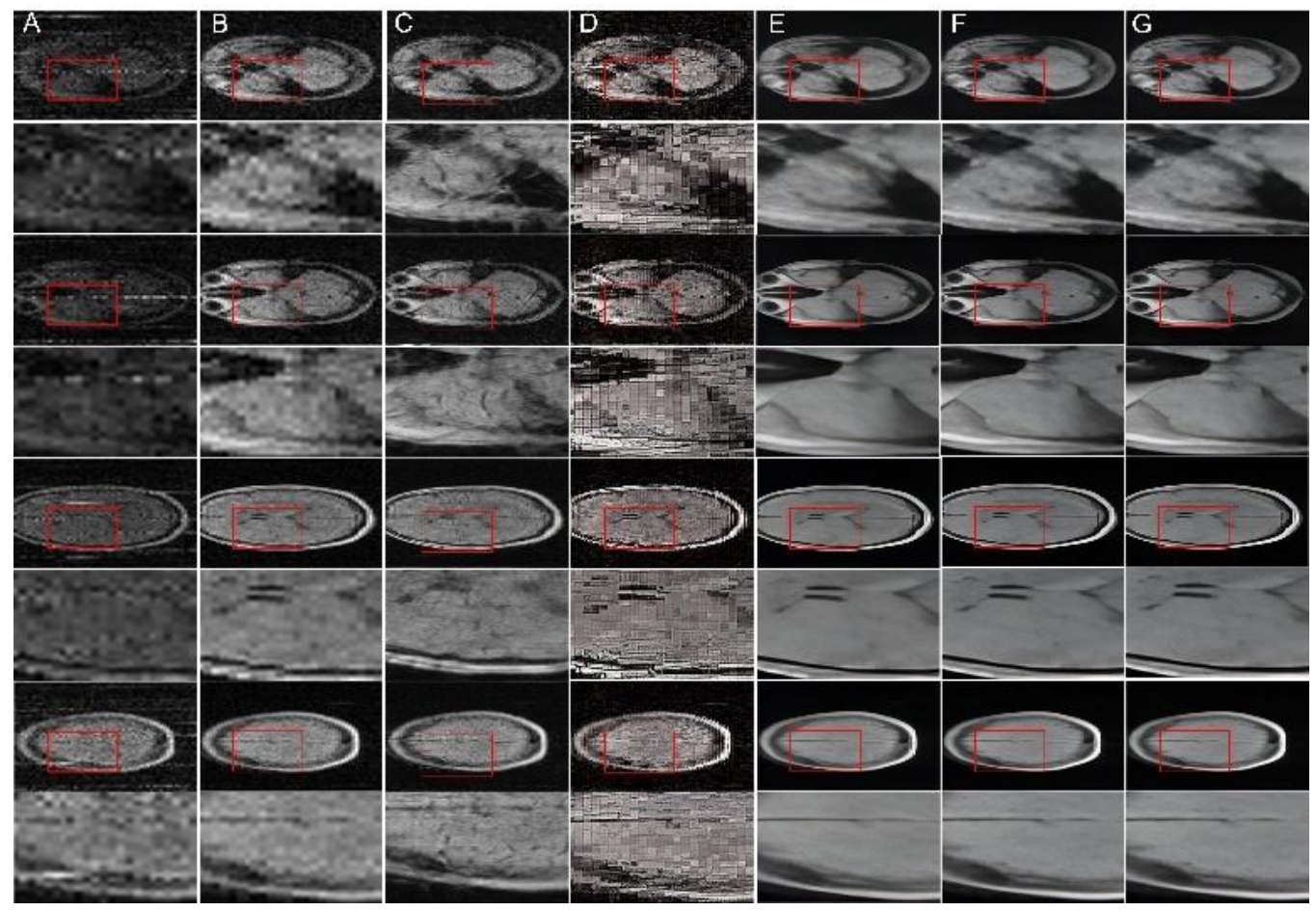

Fig. 8: Performance comparison of MR super-resolution methods. Rows 1, 3, 5, and 7 show patient 1 images, with zoomed-in sections (rows 2, 4, 6, 8) from red-boxed regions. Methods: (A) LR, (B) EDITER, (C) SRGAN, (D) ESRGAN, (E) BSRGAN, (F) Real-ESRGAN, and (G) MRSRGAN.

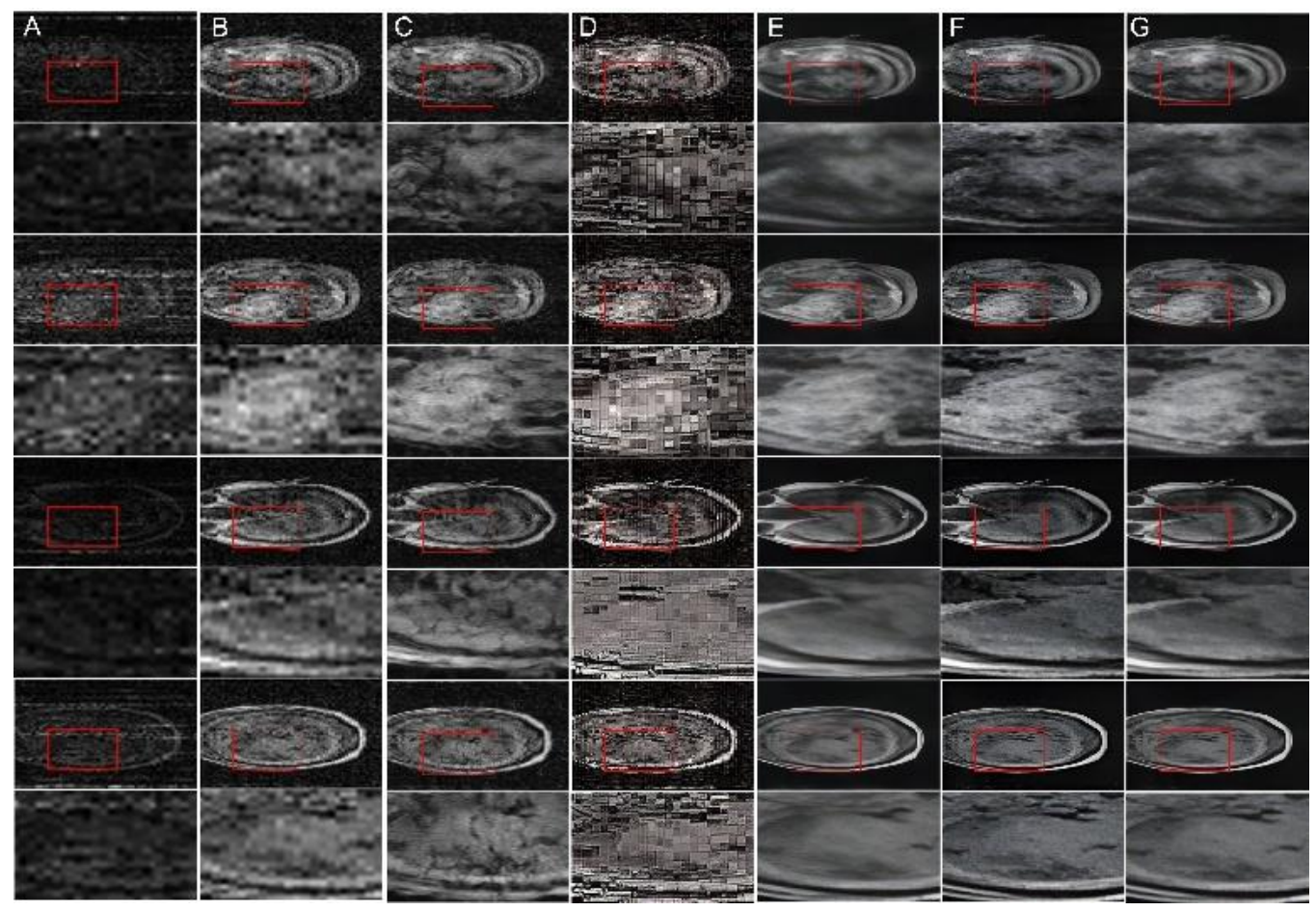

Fig. 9: Performance comparison of MR super-resolution methods. Rows 1, 3, 5, and 7 display images of patient 2, with zoomed-in sections (rows 2, 4, 6, 8) highlighting red-boxed regions. Methods: (A) LR, (B) EDITER, (C) SRGAN, (D) ESRGAN, (E) BSRGAN, (F) Real-ESRGAN, and (G) MRSRGAN.

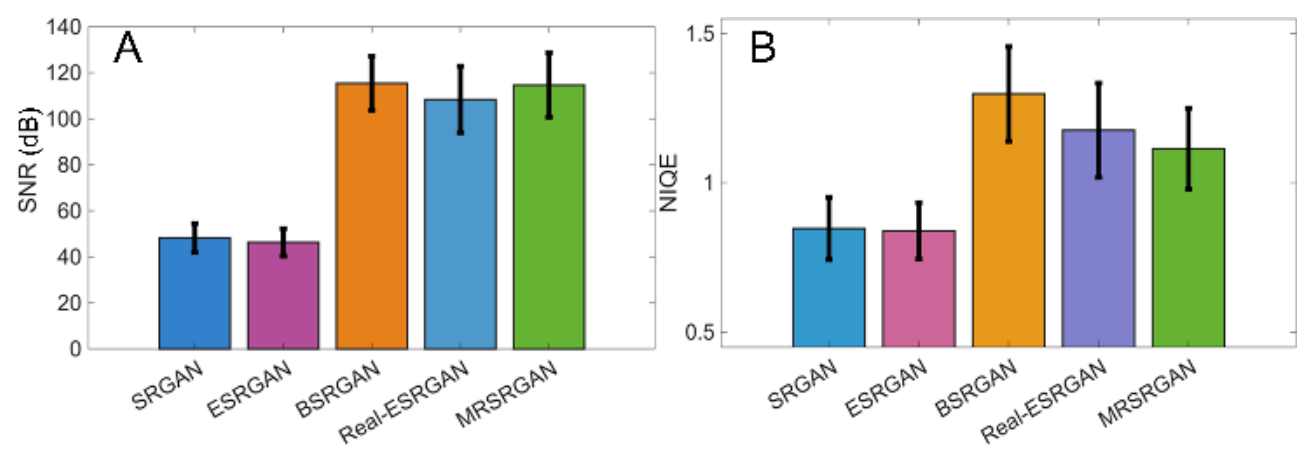

Fig. 10: Comparison of different SR methods using (A) Average SNR values (B) Average NIQE of 40 samples of our LF MRI dataset.

**List of tables on supporting information:**

**Supplementary information**

**Supporting Information Table S1:** Summary of artefacts, related parameters, and corresponding ranges considered to generate them

| Artefact | Parameter | Range |
|---|---|---|
| **Slice thickness effect ($A_1$)** | Number of slices in averaging | [2, 5] |
| **Frequency Encoding Blur($A_2$)** | Anisotropic | [12.5%, 50%] of K-space |
| **Phase Encoding Blur(A3)** | Anisotropic | [12.5%, 50%] of K-space |
| **Motion Blurring ($A_4$)** | Standard deviation $\sigma_{Gm}$ | [15, 25] |
| **Ghosting Artefact ($A_5$)** | Percentage of K-space lines to swap | $[\pi/10 , \pi]$ |
| | Rotation angle | $[0, \pi/2]$ |
| | Degree of translation | [0, h/25] [0, w/25] |
| **Chemical shift artifacts ($A_6$)** | Fat frequency shift = 3.5% ppm Bandwidth = 400 | [1, 3] |
| **Ringing Artifacts ($A_7$)** | K-space lines (Horizontal) | [0.35, 0.5] % |
| | K-space lines (Vertical) | [0.35, 0.5] % |
| **Field Inhomogeneity ($A_8$)** | Field intensity center and std dev. $\sigma_{B0}$ | [50, 30, 40, 60] |
| | Lowe bound amplitudes | 0.75*[0.3, 0.08, 0.05, 0.06] |
| | Upper bound amplitudes | [0.3, 0.08, 0.05, 0.06] |
| **Signal-Void/Shading Artifact ($A_9$)** | Distance from center | [ 0.45, 0.53] % |
| **Magnetic Susceptibility ($A_{10}$)** | Amplitude of spikes | [Mean intensity, Max intensity] |
| **Zebra Artifacts ($A_{11}$)** | Frequency | [1.5, 2.5] |
| | Amplitude | [0.3, 0.5] |
| **Down-sampling ($D_1$)** | Size (Spatial dimension) | [2, 4] |
| **Noise Degradation ($N_1$)** | Standard deviation & Rician coefficient | [0.05, 0.25] |
| **JPEG compression noise ($N_2$)** | Size reduction up to given % | [30, 90] |

**Supporting Information Table 2**: Performance comparison of various methods for chemical shift artifact-related image enhancement using a CDP dataset for variation.

| | **SRGAN** | **ESRGAN** | **BSRGAN** | **Real-ESRGAN** | **MRSRGAN** |
|---|---|---|---|---|---|
| PSNR(dB) | 26.54 | 25.62 | 29.16 | 28.13 | **30.54** |
| SSIM | 0.7129 | 0.6483 | 0.7092 | 0.7059 | **0.7332** |
| NQM | 23.22 | 21.26 | 24.87 | 25.27 | **26.12** |
| MSSSIM | 0.9034 | 0.8787 | 0.9175 | 0.9207 | **0.9476** |
| MSAM | 1.5830 | 1.5379 | 1.5462 | 1.5547 | **1.5235** |

**Supporting Information Table S3:** Performance comparison of various methods for motion artifact-related image enhancement using a SLP simulated dataset and MR-ART dataset.

| | SLP Simulated Dataset | | | | |
|---|---|---|---|---|---|
| | SRGAN | ESRGAN | BSRGAN | Real-ESRGAN | MRSRGAN |
| PSNR(dB) | 26.60 | 25.64 | **30.21** | 29.54 | 29.70 |
| SSIM | 0.6758 | 0.6282 | 0.7236 | 0.7260 | **0.7745** |
| NQM | 22.37 | 20.16 | 25.07 | 25.40 | **25.93** |
| MSSSIM | 0.9102 | 0.8401 | 0.7863 | 0.8659 | **0.9523** |
| MSAM | 1.5926 | 1.5354 | 1.5590 | 1.5631 | **1.5291** |
| | MR-ART (original) dataset | | | | |
| PSNR(dB) | 26.08 | 24.64 | **28.15** | 27.31 | 28.08 |
| SSIM | 0.6745 | 0.6501 | 0.7164 | 0.7159 | **0.7459** |
| NQM | 22.30 | 20.93 | 24.97 | 25.33 | **25.79** |
| MSSSIM | 0.9068 | 0.8698 | 0.8512 | 0.8932 | **0.9391** |
| MSAM | 1.5878 | 1.5938 | 1.5526 | 1.5589 | **1.5211** |

**Supporting Information Table S4:** Performance comparison of various methods for B0-inhomogeneity artifact-related image enhancement using a SLP simulated dataset.

| | SRGAN | ESRGAN | BSRGAN | Real-ESRGAN | MRSRGAN |
|---|---|---|---|---|---|
| PSNR (dB) | 27.84 | 26.88 | **28.58** | 27.92 | 28.28 |
| SSIM | 0.7253 | 0.7119 | 0.7456 | 0.7403 | **0.7656** |
| NQM | 26.19 | 28.52 | 29.24 | 30.31 | **31.54** |
| MSSSIM | 0.9151 | 0.8832 | 0.8975 | 0.9183 | **0.9442** |
| MSAM | 1.5206 | 1.5871 | 1.5179 | 1.5175 | **1.5103** |

**Supporting Information Table S5:** Performance comparison of various methods for MRI enhancement using simulated SLP dataset (5% noise).

| | **SRGAN** | **ESRGAN** | **BSRGAN** | **Real-ESRGAN** | **MRSRGAN** |
|---|---|---|---|---|---|
| PSNR(dB) | 28.70 | 27.20 | **31.62** | 30.88 | 31.36 |
| SSIM | 0.8112 | 0.8259 | 0.8584 | 0.8478 | **0.8792** |
| NQM | 21.45 | 25.09 | 25.04 | 25.75 | **26.03** |
| MSSSIM | 0.9134 | 0.9293 | 0.9518 | 0.9523 | **0.9793** |
| MSAM | 1.5854 | 1.5219 | 1.5494 | 1.5568 | **1.5171** |